\documentclass[12pt,a4paper]{article}
\usepackage[latin2]{inputenc}
\usepackage{amsmath}
\usepackage{amsfonts}
\usepackage{amssymb}
\usepackage{graphicx}
\usepackage{slashed}
\usepackage{braket}
\usepackage[a4paper,bindingoffset=0.2in,
            left=0.75 in,right=0.75 in,top=0.75 in,bottom=0.75  in,
            footskip=.25in]{geometry}
\usepackage{physics}

\begin{document}

\title{Connecting dissipation and noncommutativity: A Bateman system case study}  

\author{Sayan Kumar Pal\footnote{sayankpal@bose.res.in}$~^{1}$, 
Partha Nandi\footnote{parthanandi@bose.res.in}$~^{1}$ and 
Biswajit Chakraborty\footnote{biswajit@bose.res.in}$~^{1}$\\
{\small{$^{1}$S. N. Bose National Centre for Basic Sciences}}\\
{\small{JD Block, Sector III, Salt Lake City, Kolkata 700106, India}}}
\maketitle
\begin{abstract}
Quantum effects on a pair of Bateman oscillators embedded in an ambient noncommutative space (Moyal plane) is analyzed using both path integral and canonical quantization schemes within the framework of Hilbert-Schmidt operator formulation. We adopt a method which is distinct from the one which employs 't Hooft's scheme of quantization, carried out earlier in the literature where the ambient space was taken to be commutative. Our quantization shows that we end up finally again with a Bateman system except that the damping factor undergoes renormalization. The corresponding expression shows that the renormalized damping factor can be non-zero even if  "bare" one is zero to begin with. Conversely, the noncommuatative parameter $\theta$, taken to be a free one now, can be fine-tuned to get a vanishing renormalized damping factor. This indicates a duality between dissipative commutative theory and non-dissipative noncommutative theory.
\end{abstract}

\section{Introduction} 
It was long time back in 1930's that H. Bateman \cite{Bat} formulated a time indepedent Lagrangian analysis for the damped harmonic oscillator (DHO) problem by complementing the DHO by its time reversed image and worked with an effective doubled system. The latter system corresponds to the so-called "anti-damped" oscillator. The corresponding pair of equations are given by -

\begin{equation}
\ddot{x} + \gamma\dot{x} + \omega^2x = 0\label{d1}
\end{equation}
and,
\begin{equation}
\ddot{y} - \gamma\dot{y} + \omega^2y = 0 \label{adamped}
\end{equation}
and either of these equations can be obtained from the following action - 
\begin{equation}
S=\int dt L~~~;~~L = \dot{x}\dot{y} + \frac{\gamma}{2}\left(x\dot{y} - \dot{x}y\right)
       - \omega^2xy\label{n}
\end{equation}
Here the parameters $\gamma>0$ and $\omega$ are independent of time representing the damping parameter and the frequency respectively. Also the mass m has been taken to be unity. On taking the following ansatz for (\ref{d1}), $x(t)=e^{i\lambda t}$, one finds that $\lambda$ is given by $\lambda_{\pm}= i\frac{\gamma}{2} \pm \frac{\gamma}{2}\sqrt{R-1} \label{fre}$, where $R =\frac{4\omega^2}{\gamma^2}$ is
the critical ratio. If $R>1$, the motion for the system described by (\ref{d1}) is oscillatory with
exponentially decaying amplitude. Otherwise, the motion is nonoscillatory and overdamped.

In modern day literature, this form (\ref{n}) is known as the indirect representation \cite{Morse},\cite{Sant}.
This represents a closed system with a damped (source) and anti-damped (sink) harmonic oscillators along x and y axes on the 2D real plane.
It is important to mention here in this context that any of this above pair of equations cannot be described individually through a time-independent Lagrangian\footnote{A time-dependent Lagrangian $L=e^{-\gamma t}(\frac{1}{2}{\dot{x}}^2-\frac{1}{2}\omega^2 x^2)$ was proposed in \cite{Eds},\cite{Roy} for (\ref{d1}).} ~ as they do not correspond to closed classical systems. Although a time-independent Lagrangian (\ref{n})  can be written down, the quantization of such a system remains a non-trivial task. And this stems from the fact that the Legendre-transformed Hamiltonian turns out to have no lower bound(see section 2). Quantization of such systems in some cases like this, can however be carried out by adopting 't Hooft's scheme \cite{hoo}, as shown in \cite{Vit2} by defining a suitable physical subspace where the Hamiltonian has a lower bound. It turns out that quite a few interesting physical models can be re-casted in a form, so that 't Hooft's scheme of quantization (of which we provide a brief review in the Appendix) can be adapted. In fact the well-known Gupta-Bleuler quantization of free Maxwell theory can also be casted in a form similar to 't Hooft's scheme \cite{vit18}. Indeed this scheme of quantization was proposed in \cite{hoo} as an attempt to understand Planck scale physics (with associated length scale $l_p=\sqrt{\frac{\hbar G}{c^3}}\approx 10^{-33}~ cm$ ), where the existence of a possible underlying relation between quantization and dissipation of information was pointed out. Further, the role of dissipation at Planck scale was inferred from the fact that black holes, absorbing informations are inevitably formed in general relativity. In fact the ratio of coefficient of viscosity to mass density having a dimension of length scale (in natural units) was speculated to be of the order of Planck length, so that one may speculate that the physics of strictly continuous fields below Planck length scale is entirely controlled by viscosity itself. Further in certain situations, this scheme of quantization can imply implicitly a noncommutative algebra between the coordinates within this physical subspace(see Appendix). 

On the other hand, very strong plausibility arguments were provided by \emph{Doplicher et. al.} \cite{Dop} that the gravitational collapse through the formation of a black hole and arising in the process of localization of an event down to Planck length scale can perhaps be avoided in a noncommutative (NC) spacetime where one postulates a NC algebra between spacetime coordinates which are now elevated to the level of operators. In fact, such expectations were indeed shown to be true by considering the system of a simple harmonic oscillator in 2D Moyal plane ($[\hat{x_1},\hat{x_2}]=i\theta$) \cite{Sir} and also in 1+1 Moyal spacetime ($[\hat{t},\hat{x}]=i\theta$) \cite{Nand}. In either case, the spread of ground state wave function or for that matter that of the probability density was shown to be greater than the natural length scale $\sqrt{\theta}$ and cannot be squeezed below this scale even in the limit of infinite confining potential. In this scenario, therefore NC can play a role in preventing the formation of a black hole. Besides being the simplest among all other possible forms of the NC algebra, the Moyal spaces/spacetimes have become most popular toy model to investigate quantum gravity effects, as this can be taken as a prototype spacetime at this scale.

In the present paper, we are therefore motivated to study the effect of noncommutativity at the quantum level by placing this system described by ({\ref{n}}) in an ambient noncommutative(NC) space and quantize the system employing both path integral and canonical quantization, which is distinct and inequivalent to the above mentioned 't Hooft scheme. Consequently there will be no noncommutativity which is "internally" generated, rather it is imposed externally by hand, through the ambient NC space. Particularly we will be interested to study the NC quantum effects arising from NC coordinate algebra satisfied by the coordinates x and y appropriate for the Moyal plane. Apart from its simple structure, there is already a huge literature in this area, both at the quantum mechanics and quantum field theoretical level. Further, most of these analyses employed Moyal or Voros star product. However, it was noted in \cite{lizzi} that the formulations using Moyal or Voros star products are not necessarily equivalent. This problem can be bypassed in a Hilbert-Schmidt (HS) operatorial formulation, initiated in \cite{Sir}, where one confronts operatorial nature of the space and/or time variables head on. At best, one can make use of the Voros product or the associated basis (see section 2.2). In this present paper also we shall be making use of this HS operatorial formulation.

The paper is organized as follows. In the next section, we shall provide the necessary background materials, required for this paper and formulate the problem where we augment the Lagrangian with planar isotropic oscillator involving suitable pair of parameters representing mass and angular frequency, apart from placing the system in an ambient NC space-taken to be the Moyal plane. We then carry out the path integral and canonical quantization of the problem in section 3. In section 4, we study the possibility of taking vanishing limits for both of these parameters , so that one has a pair of pure Bateman oscillators eventually. We finally conclude in section 5.
\section{}
\subsection{Review of the background materials and setting up of the problem}
Here, in this sub-section, we would like to provide a brief review of the schemes adopted to quantize the system, described by the Lagrangian (\ref{n}) and then introduce the scheme we are going to adopt. We begin by recapitulating the Bateman method \cite{Bat} to obtain (\ref{n}), in the so-called "indirect representation"~.
To that end, let us look for an action $S$ whose infinitesimal variation has the following structure 
\begin{equation}
\delta S = - \int^{t_2}_{t_1}dt\left[\left(\ddot{x} 
         +\gamma\dot{x} + \omega^2 x\right)\delta y +
                \left(\ddot{y} - \gamma\dot{y} + \omega^2 y \right)\delta x\right]
                          \label{batf}
\end{equation}
so that variation w.r.t. x and y with the usual boundary conditions where $\delta x$ and $\delta y$ vanishes at $t_1$ and $t_2$ automatically yields (\ref{adamped}) and (\ref{d1}) respectively.
On integrating by parts and making use of the usual boundary conditions, it is possible to bring it to an integrable form given by-
\begin{equation}
\delta S = \delta \int^{t_2}_{t_1}dt\left[\dot{x}\dot{y} + \frac{\gamma}{2}\left(x\dot{y} - \dot{x}y\right)
       - \omega^2xy \right]
                          \label{batf2}
\end{equation}
so that the action $S$ (\ref{n}) can be read off easily from the above equation. Now, introducing the transformed coordinates $x_1$ and $x_2$ where,
\begin{equation}
	\begin{pmatrix}
	x \\
	y
	\end{pmatrix} \longrightarrow
	\begin{pmatrix}
	x_{1} \\
	x_{2} 
	\end{pmatrix}
	=T_1 	\begin{pmatrix}
		x \\
		y
		\end{pmatrix}
	;~ T_1 = \frac{1}{\sqrt{2}}\begin{pmatrix}
	1 & 1 \\
	1 & -1 \
	\end{pmatrix} \label{t1}
	\end{equation}
the Lagrangian (\ref{n})  can be written in a compact notation as 
\begin{equation}
L = {1\over 2}g_{ij}\dot{x}_i\dot{ x}_j - {\gamma\over 2}\epsilon_{ij}x_i\dot{x_j} - {\omega^2\over 2}g_{ij}x_i x_j \label{Lc}
\end{equation}
\footnote{Except for this Lorentzian signature, the composite Lagrangian (\ref{Lc}) has essentially the same structural form as that of the general two dimensional Euclidean planar oscillator in a magnetic field where $g_{ij}=\delta_{ij}$ and $\gamma$ plays the role of magnetic field.}where  $g_{ij}$ is the pseudo - Euclidean i.e. Lorentzian metric: $g_{11}$ = -$g_{22}$ = 1 and $g_{12}$ = 0.

 The Hamiltonian corresponding to (\ref{Lc}) is given by-
\begin{equation}
H = {1\over 2}(p_1-\frac{\gamma x_2}{2})^2- {1\over 2}(p_2+\frac{\gamma x_1}{2})^2 + {1\over 2} \omega^2 (x_1^2 -x_2^2) \label{L4}
\end{equation}

and can be written as a difference of two positive hamiltonians $H_1$ and $H_2$ -
\begin{eqnarray}\label{hsplit}
 H = H_1-H_2~;~ H_1={1\over 2}(p_1-\frac{\gamma x_2}{2})^2 + {1\over 2} \omega^2 x_1^2 ~, and ~H_2= {1\over 2}(p_2+\frac{\gamma x_1}{2})^2  + {1\over 2} \omega^2 x_2^2 
\end{eqnarray}

indicating that it is not bounded from below.
 Quantization of the Bateman system in the spirit of 't Hooft \cite{hoo} has been carried out by \emph{Blasone et. al.} \cite{Vit2} . However we take a different and inequivalent approach in this paper where we include additional interactions to (\ref{d1}) and (\ref{adamped}) to render the Hamiltonian positive and this will be achieved through the introduction of linear and second-order derivative couplings between x and y oscillators for which the strength parameters are $\epsilon$ and $\eta$ respectively. Thus instead of (\ref{d1}) and (\ref{adamped}) we have the following pair of equations - 
\begin{equation}
\ddot{x} + \gamma\dot{x} + \omega^2x = -\epsilon y - \eta \ddot{y}\label{G1}
\end{equation}
\begin{equation}
\ddot{y} -\gamma\dot{y} + \omega^2y = -\epsilon x - \eta \ddot{x}\label{G2}
\end{equation}
which in turn follow from the Lagrangian :
\begin{equation}
L = \dot{x}\dot{y} + \frac{\gamma}{2}\left(x\dot{y} - \dot{x}y\right)- \omega^2xy + \frac{\eta}{2}(\dot{x^2}+\dot{y^2}) - \frac{\epsilon}{2}(x^2+y^2)\label{LE}
\end{equation}

This represents the Lagrangian of a damped Bateman oscillator augmented by a pure 2d harmonic oscillator Lagrangian with mass $\eta$ and spring constant $\epsilon $~.
Hamiltonian analysis of Bateman oscillator was constructed by Morse, Feshback and Tikochinsky \cite{Fesh}, \cite{Tik}, \cite{Morse} and subsequently  Lemos \cite{Lem} studied the associated Hamilton-Jacobi formulation.
A very good account of the classical and quantum mechanics of the damped harmonic oscillator was provided in \cite{Dekk}. In \cite{rabin}, a canonical approach to dissipation was analyzed in terms of the elementary chiral modes of the problem. Recently, Bender et al \cite{Bend} have considered an interacting case in which the two modes of the Bateman system was coupled linearly. Through their mathematical model, they have explained a system of two coupled whispering-gallery-mode optical resonators on which currently many experiments \cite{nat} have been carried out. The introduction of second-order derivative couplings along with a linear coupling is completely new that we have been addressing here for the first time in the literature to best we are aware of. It is to be mentioned here that both of these couplings will play a crucial role in our quantum mechanical formulation which will become transparent towards the end.



In terms of ($x_1,x_2$) coordinates (\ref{t1}), the "master equations" (\ref{G1}) and (\ref{G2}) take the following forms-
\begin{equation}
(\eta+1)\ddot{x_1} + \gamma\dot{x_2} + (\epsilon + \omega^2)x_1 = 0
\end{equation}
\begin{equation}
(\eta-1)\ddot{x_2} -\gamma\dot{x_1} + (\epsilon - \omega^2)x_2 = 0
\end{equation}
while the Lagrangian (\ref{LE}) takes the form-
\begin {equation}
L = {(\eta+1)\over 2}\dot{x_1^2}+{(\eta-1)\over 2}\dot{x_2^2} - \frac{\gamma}{2}\left(x_1\dot{x_2}-x_2\dot{x_1}\right) -  {(\epsilon+\omega^2)\over 2} x_1^2-{(\epsilon-\omega^2)\over 2}x_2^2\label{G3}
\end{equation}
which can again be re-casted in a compact form like (\ref{Lc}) as,
\begin{equation}
L = {1\over 2}\tilde{g_{ij}}\dot{x}_i\dot{ x}_j - {\gamma\over 2}\epsilon_{ij}x_i\dot{x_j} - {\tilde{\omega_i}^2\over 2}x_i^2\label{L3}
\end{equation}
where the metric  $\tilde{g_{ij}}$ is now given by $\tilde{g_{11}} = \eta+1,~ \tilde{g_{22}} =\eta- 1$ and $ \tilde{g_{12}}$ = 0.
Besides, $\tilde{\omega_1}^2=\epsilon+\omega^2$ and $\tilde{\omega_2}^2=\epsilon-\omega^2$, so that it is no longer isotropic.

And the corresponding Hamiltonian is 
\begin{equation}
\text{\footnotesize $H = \frac{p_1^2}{2(\eta+1)}+\frac{p_2^2}{2(\eta-1)}+\frac{\gamma}{2}\left(\frac{x_1p_2}{\eta-1}-\frac{x_2p_1}{\eta+1}\right)+\left(\frac{\gamma^2}{8(\eta-1)}+ {(\epsilon+\omega^2)\over 2}\right) x_1^2+\left(\frac{\gamma^2}{8(\eta+1)}+ {(\epsilon-\omega^2)\over 2}\right) x_2^2 $} \label{ha}
\end{equation}
The positive definiteness of the Hamiltonian can now be ensured if we demand  $\eta > 1$  and $\epsilon > \omega^2$.

This form of the Hamiltonian is still not well suited for carrying on the path-integral formulation of the problem because of different effective masses for the two modes. To convert it into a useful form, we now implement a second canonical transformation $T_2$ given by:
\begin{eqnarray}
&(x_1,x_2,p_1,p_2)^{T} \longrightarrow (x_1',x_2',p_1',p_2') ^{T}=T_2(x_1,x_2,p_1,p_2)^{T}  \\
&where,~~~~~ T_2=diag\bigg((\frac{\eta + 1}{\eta - 1})^\frac{1}{4},~ (\frac{\eta - 1}{\eta + 1})^\frac{1}{4}, ~(\frac{\eta - 1}{\eta + 1})^\frac{1}{4}, ~(\frac{\eta + 1}{\eta - 1})^\frac{1}{4}\bigg).  ~~~~~~~~~~~~~~~~~~~~~~\label{t2}
\end{eqnarray}
so that in terms of the transformed primed variables, the Hamiltonian can be rewritten as,
\begin{equation}
H = \frac{p_1'^2}{2\mu}+\frac{p_2'^2}{2\mu}+\frac{\gamma}{2\mu}(x_1'p_2'-x_2'p_1')+\frac{1}{2}\mu\omega_1^2 x_1'^2+\frac{1}{2}\mu\omega_2^2 x_2'^2 \label{Final}
\end{equation}
where the frequencies are given by
\begin{equation} \label{frequen}
 \omega_1^2=\frac{\gamma^2}{4(\eta^2-1)}+\frac{\epsilon+\omega^2}{\eta+1}~~,~\omega_2^2=\frac{\gamma^2}{4(\eta^2-1)}+\frac{\epsilon-\omega^2}{\eta-1}~~.
 \end{equation}
 Here $\mu= \sqrt{(\eta+1)(\eta-1)}$ can be regarded as the new mass parameter and is given by the geometric mean of the two masses ($\eta+1$) and ($\eta-1$) occurring in (\ref{ha}).
\\~~~~~~~~~~~~~~~~~~~~~~~~~~~~~~~~~~~~~~~~~~~~~~~~~~~~~~~~~~~~~~~~~~~~~~~~~~~~~~~~~~~~~~~~~~~~~~~~~~~~~~~~~~~~~~~~~~~~~~~~~~~~~~~~~~~~~~~~~~~~~~~~~~~~~~
\\
This will be the essential form of the Hamiltonian with which we are going to work in sections to follow except that the system is now placed in the ambient noncommutative Moyal plane, where $\hat{x_1'}$ and $\hat{x_2'}$ now satisfy the following noncommutative algebra:
\begin{equation}
[\hat{x}_1', \hat{x}_2']= [\hat{x}_1, \hat{x}_2] =  [\hat{y}, \hat{x}]= i\theta \label{algebra}
\end{equation}

Note that in our notation, it is $[\hat{y}, \hat{x}]=i\theta$, rather than $[\hat{x}, \hat{y}]=i\theta$. This is because $det~ T_1=-1 ~~(\ref{t1})$.
\\

In what follows we shall carry out the quantization of this system in both path integral and canonical quantization schemes in the Hilbert-Schmidt operatorial formulation, introduced in \cite{Sir},\cite{Schol}, a brief review of which is provided below.




\subsection{Review of Hilbert-Schmidt operator formulation}

\noindent 
The formal and interpretational aspects of noncommutative quantum mechanics was introduced in \cite{Sir} and \cite{Schol}. It can be viewed as a quantum system represented in the space of Hilbert-Schmidt operators acting on noncommutative configuration space identified as the Hilbert space furnishing a representation of just the coordinate algebra (\ref{algebra}) . Based on this formalism, the path integral formulation and derivation of the action for a free particle moving in a noncommutative plane was done in \cite{Sun} using coherent states. 



The Moyal plane, defined through (\ref{algebra}) has\footnote{From now on, we shall be essentially working with the primed coordinates $x_1',  x_2'$ (\ref{t2}) but for brevity the primes will be omitted. We will revert back to ($x_1,x_2$) and (x,y) coordinates by considering inverse transformations $T_2^{-1}$ (\ref{t2}) and  $T_1^{-1}$ (\ref{t1}) only in section 5.} ~~four-dimensional phase space where the associated operators follow NC Heisenberg algebra
\begin{eqnarray}
\label{NHA}
\left[\hat{x}_{i},\hat{x}_{j}\right]=i\theta_{ij} = i\theta \epsilon_{ij} ~ ; \quad \left[\hat{p}_{i},\hat{p}_{j}\right]=0~ ; \quad \left[\hat{x}_{i},\hat{p}_{j}\right] = i{\hbar}\delta_{ij}.
\end{eqnarray}
Here $\theta$ denotes the constant spatial noncommutative parameter that can be taken to be positive ($\theta>0$) without loss of generality.


In order to construct the NC configuration space, let us first observe that the coordinate algebra (\ref{algebra}) is isomorphic to the 1-D harmonic oscillator. We can therefore identify our NC configuration space Hilbert space to be the same as that of 1-D harmonic oscillator. In order to introduce it formally, let us define the annihilation and creation operators as
$\hat b = \frac{1}{\sqrt{2\theta}} (\hat{x}_1+i\hat{x}_2)$,
$\hat{b}^\dagger =\frac{1}{\sqrt{2\theta}} (\hat{x}_1-i\hat{x}_2)$
satisfying the Fock algebra $[\hat{b}, \hat{b}^\dagger ] = 1$. 
The noncommutative configuration space is then 
isomorphic to the bosonic Fock space
\begin{equation}
\mathcal{H}_c = \textrm{span}\{ |n\rangle= 
\frac{1}{\sqrt{n!}}(\hat{b}^\dagger)^n |0\rangle\}_{n=0}^{n=\infty}
\label{Fock}
\end{equation}
This $\mathcal{H}_c $ furnishes the representation of the coordinate algebra (\ref{algebra}). However, to do quantum mechanics on NC Moyal plane we need a Hilbert space, which furnishes a representation of the entire NC Heisenberg algebra (\ref{NHA}) which contains (\ref{algebra}) as a sub-algebra. And this corresponds to the space $\mathcal{H}_q $ of Hilbert-Schmidt (HS) operators i.e. the space of composite operators $\psi(\hat{x}_1, \hat{x}_2)$, built out of $\hat{x}_1$ and $ \hat{x}_2$  having a finite HS norm
\begin{equation}
\mathcal{H}_q = \left\{ \psi(\hat{x}_1,\hat{x}_2): 
\psi(\hat{x}_1,\hat{x}_2)\in \mathcal{B}
\left(\mathcal{H}_c\right),\;
{\rm tr_c}(\psi^\dagger(\hat{x}_1,\hat{x}_2)
\psi(\hat{x}_1,\hat{x}_2)) < \infty \right\}.
\label{4}
\end{equation}
Here ${\rm tr_c}$ denotes tracing over  $\mathcal{H}_c $  and $\mathcal{B}\left(\mathcal{H}_c\right)$ 
the set of bounded operators on $\mathcal{H}_c$. 
This space has a natural inner product-
\begin{equation}
\left(\phi(\hat{x}_1, \hat{x}_2), \psi(\hat{x}_1,\hat{x}_2)\right) = 
{\rm tr_c}(\phi(\hat{x}_1, \hat{x}_2)^\dagger\psi(\hat{x}_1, \hat{x}_2))< \infty
\label{inner}
\end{equation}
and stems from the fact that the product of two HS operators has finite trace-class norm. Also,  $\mathcal{H}_q $ forms a two-sided *-ideal in the Banach algebra $\mathcal{B}\left(\mathcal{H}_c\right)$  and forms a Hilbert space on its own. In fact, one can identify $\mathcal{H}_q$ as $\mathcal{H}_c \otimes \mathcal{H}_c^*$, with $\mathcal{H}_c^*$ being the dual of $\mathcal{H}_c$ (\ref{Fock}).
To distinguish between vectors belonging to $\mathcal{H}_c$ and $\mathcal{H}_q$, we use angular $(|.>)$ and round $(|.))$ kets respectively. An unitary representation of the noncommutative Heisenberg algebra (\ref{NHA}) is obtained through the following actions of $\hat{X_i}$'s and $\hat{P_i}$'s on $\mathcal{H}_q$ :
\begin{equation} 
\hat{X}_i \psi (\hat{x}_1, \hat{x}_2) = \hat{x}_i \psi (\hat{x}_1, \hat{x}_2)~,~ \hat{P}_i \psi (\hat{x}_1, \hat{x}_2) = \frac{\hbar}{\theta}\epsilon_{ij} [\hat{x}_j, \psi (\hat{x}_1, \hat{x}_2)] \label{act}
\end{equation}
Note that both $\hat{X_i}$'s and $\hat{x_i}$'s satisfy same i.e. isomorphic algebra and are essentially the same, except that their domains of actions are different, i.e. $\mathcal{H}_q $ and $\mathcal{H}_c $ respectively. One can regard  $\hat{X_i}$ as the representation of $\hat{x_i}$.
In view of the absence of common eigenstates of $\hat{x_1}$ and $\hat{x_2}$, the best one can do is to introduce the minimal uncertainty states i.e. maximally localized coherent states on $\mathcal{H}_c$ as-
\begin{equation}
\label{cs} 
|z\rangle = e^{-z\bar{z}/2}e^{z b^{\dagger}} |0\rangle ~\in ~\mathcal{H}_c~~~~~;~~
\Delta \hat{x}_1 \Delta \hat{x}_2= \frac{1}{2}\theta
\end{equation}
where, $z=\frac{1}{\sqrt{2\theta}}\left(x_1+ix_2\right)$ 
is a dimensionless complex number. Corresponding to these states we can construct a state (HS operator) in $\mathcal{H}_q$ as follows-
\begin{equation}
|z, \bar{z} ) \equiv|\vec{x}) \equiv|x_1, x_2 )=\frac{1}{\sqrt{2\pi\theta}}|z\rangle\langle z| \in \mathcal{H}_q~~;~~~ (z^\prime, \bar z^\prime|z, \bar z)=e^{-|z-z^\prime|^2} \label{base}
\end{equation}
These states, which we refer to as Voros basis in the sequel for its association with Voros star product (see (\ref{starp}) below), have the property
\begin{equation}
\hat{B}|z, \bar{z})=z|z, \bar{z})~;~\hat{B}=\frac{1}{\sqrt{2\theta}}(\hat{X_1}+i\hat{X_2}).
\label{p1}
\end{equation}
Again $\hat{B}$ can be regarded ss a representation of $\hat{b}$.
The `position' representation, or the so-called 'symbol' of a state 
$|\psi)=\psi(\hat{x_1},\hat{x_2})$ can be constructed by taking overlap with (\ref{base}) as,
\begin{equation}
(x_1, x_2|\psi)\equiv(z, \bar{z}|\psi)=\frac{1}{\sqrt {2\pi\theta}}tr_{c}
(|z\rangle\langle z| \psi(\hat{x}_1,\hat{x}_2))=
\frac{1}{\sqrt {2\pi\theta}}\langle z|\psi(\hat{x}_1,\hat{x}_2)|z\rangle.
\label{symb}
\end{equation}
We now introduce the momentum eigenstates as-
\begin{equation}
|p)=\sqrt{\frac{\theta}{2\pi\hbar^{2}}}e^{\frac{i}{\hbar}\vec{p}.\hat{\vec{x}}}=\sqrt{\frac{\theta}{2\pi\hbar^{2}}}e^{i\sqrt{\frac{\theta}{2\hbar^2}}
(\bar{p}b+pb^\dagger)}~~;~\hat{P}_i |p)=p_i |p)~; ~~p=p_1+ip_2
\label{eg}
\end{equation}
satisfying the completeness and orthonormality relations-
\begin{eqnarray}
\int d^{2}p~|p)(p|=1_{Q}~. ~~;~~ (\vec{p'}|\vec{p})=\delta^2(\vec{p'}-\vec{p})
\label{eg5}
\end{eqnarray}
We now observe that the wave-function of a "free particle" on the noncommutative plane is given by
\begin{eqnarray}
(z, \bar{z}|p)=\frac{1}{\sqrt{2\pi\hbar^{2}}}
e^{-\frac{\theta}{4\hbar^{2}}\bar{p}p}
e^{i\sqrt{\frac{\theta}{2\hbar^{2}}}(p\bar{z}+\bar{p}z)}~.
\label{eg3}
\end{eqnarray}
The completeness relations for the position eigenstates $|z,\bar{z})$ (which is an important
ingredient in the construction of the path integral representation) reads
\begin{eqnarray}
\int 2\theta dzd\bar{z}~|z, \bar{z})\star(z, \bar{z}|=\int dx_
1 dx_2~|x_1, x_2)\star(x_1, x_2|=1_{Q}
\label{eg6}
\end{eqnarray}
where the Voros star product between two functions 
$f(z, \bar{z})$ and $g(z, \bar{z})$ is defined as
\begin{eqnarray}
f(z, \bar{z})\star g(z, \bar{z})=f(z, \bar{z})
e^{\stackrel{\leftarrow}{\partial_{\bar{z}}}
\stackrel{\rightarrow}{\partial_z}} g(z, \bar{z})~.
\label{eg7}
\end{eqnarray}
This Voros star product enables us to establish an isomorphism  \cite{Basu} between $\mathcal{H}_q$ (-regarded as an algebra through the natural product of operators) and that of the space of symbols, provided that the pair belonging to the latter space is composed through Voros star product.
\begin{equation}\label{starp}
(z, \bar{z}|\psi\phi)=(z, \bar{z}|\psi)*(z, \bar{z}|\phi)~~;~~ \forall~ \psi,\phi ~\in ~\mathcal{H}_q
\end{equation}
\\
In this context, we would like to point out that this Voros star product has a non-trivial relationship with the well-known Moyal star product which is also related to a similar isomorphism between operators and corresponding symbols. This fact has a bearing on our formulation to be presented in the sequel. We therefore conclude this section with a brief review of the same.

\paragraphmark ~To this end, first observe that the action of $\hat{X_i's}$ of $\mathcal{H}_q$ has been taken as a left action in (\ref{act}) by default. One can likewise introduce a right action as,
\begin{equation}
\hat{X_i}^R \psi(\hat{x}_1,\hat{x}_2)=\psi(\hat{x}_1,\hat{x}_2)\hat{x}_i
\end{equation}
and then take their average to get the a commuting pair of variables as
\begin{equation} 
\hat{X}_i^c = \frac{\hat{X}_i^L + \hat{X}_i^R}{2} ~;~ \left[ \hat{X_1}^c, \hat{X_2}^c \right] = 0 \label{cv}
\end{equation}
A common set of eigenstates of $\hat{X_1}^c$ and $ \hat{X_2}^c$, called Moyal basis, can then be defined as-

\begin{equation}
|x)_M=\int \frac{d^2p}{(2\pi)^{2}}e^{-ip\cdot x}|p)=\sqrt{\frac{\theta}{2\pi\hbar^2}}\int\frac{d^2p}{(2\pi)^2}e^{ip\cdot(\hat x-x)} ~~ \in \mathcal{H}_q
\end{equation}
satisfying
\begin{equation}
\hat{X_i}^c|x)_M=x_i|x)_M ~~~; ~~_M(\vec{x}|\vec{y})_M=\delta^2(\vec{x}-\vec{y})
\end{equation}

One can then construct Moyal symbols as $_M(\vec{x}|\psi)$ and then an isomorphism between operators belonging to $\mathcal{H}_q$ and these symbols can be established provided the latter elements are composed through Moyal star product $*_M $ :
\begin{equation}
_M(\vec{x}|\psi\phi)=~ _M(\vec{x}|\psi)*_M ~ _M(\vec{x}|\phi)~~;~~*_M=e^{\frac{i}{2}\theta \epsilon_{ij} \stackrel{\leftarrow}{\partial_{i}}
\stackrel{\rightarrow}{\partial_{j}}}
\end{equation}
However, as was explained in \cite{Basu} , these Moyal basis and the associated Moyal star product cannot be regarded as physical; they are basically mathematical constructs for the following reasons -\\
\textbf{.} Moyal basis do not conform to POVM (positive operator-valued measure) and therefore not amenable to probabilistic interpretation of NC quantum mechanics. This is in contrast with the Voros basis (\ref{base}).\\
\\
\textbf{.} Since $\hat{X}_i^c$'s are commutative (\ref{cv}), they cannot capture the noncommutative aspects of the theory. They are unphysical position-like variables. In fact, the linear transformation:
\begin{equation}\label{cv2}
\hat{X_i}=\hat{X_i}^c-\frac{\theta}{2\hbar}\epsilon_{ij}\hat{P}_j
\end{equation}
connecting the phase space operators $(\hat{X_i},\hat{P_i})\longrightarrow (\hat{X_i}^c,\hat{P_i})$, as obtained by making use of (\ref{act},~\ref{cv}) is not a canonical one. Consequently, the physical implications of the theory where $\hat{X}_i^c$'s are regarded as physical will be different and inequivalent to the one where $\hat{X}_i$'s are regarded as physical. \\
This becomes more transparent from the fact that the symbols (\ref{symb}) in Voros basis $|\vec{x})$~(\ref{base}) and that of Moyal basis $|\vec{x})_M$ are connected by a \emph{non-invertible} map $T$ as-
\begin{equation}
(\vec{x}|\psi)=T~ _M(\vec{x}|\psi) ~~~;~~T= e^{\frac{\theta}{4}\grad^2}
\end{equation}
and this is related to the fact that the Moyal symbols belong to the Schwartz class of functions \cite{Szab}, where only the asymptotic properties of the functions and derivatives are constrained suitably, with no specifications in the ultra-violet limit i.e. in the vicinity of the small length scale $\approx{\sqrt{\theta}}$. In contrast, the Voros symbols are smooth at this tiny length scale, where the wavelengths $\leq \sqrt{\theta}$ gets automatically suppressed exponentially. Consequently, the transition matrix elements, computed in a path integral approach will depend non-trivially on whether one works in Moyal or Voros basis. Clearly, it is only the Voros basis which can provide a physical transition matrix element. We are therefore motivated to carry out the path integral quantization of the system (\ref{Final}) in the Voros basis and compute the effective action, which incorporates the quantum corrections. We take up this in the next section.

\section{}
\subsection{Path integral quantization}
With the formalism discussed in the previous section and the completeness relations for the
momentum and the position eigenstates 
(\ref{eg5}, \ref{eg6}) in place, we now proceed to write down the path integral for the propagation kernel on the two dimensional noncommutative plane. This reads
\begin{eqnarray}
(z_f, t_f|z_0, t_0)&=&\lim_{n\rightarrow\infty}\int
\prod_{j=1}^{n}(dz_{j}d\bar{z}_{j})~(z_f, t_f|z_n, t_n)\star_n
(z_n, t_n|....|z_1, t_1)\star_1(z_1, t_1|z_0, t_0)~.
\label{pint1}
\end{eqnarray}

The Hamiltonian (\ref{Final}) acting on the quantum Hilbert space $\mathcal{H}_q$ now reads
\begin{eqnarray}
\hat{H}=\frac{\hat{\vec{P}}^2}{2\mu}+ \frac{\gamma}{2\mu}(\hat{X_1}\hat{P}_2 -\hat{X_2}\hat{P}_1) +\frac{1}{2}\mu(\omega_1^2\hat{X_1}^2 +\omega_2^2\hat{X_2}^2)
\label{hami}
\end{eqnarray} 

With this Hamiltonian, we now compute the transition matrix element over a small segment in the
above path integral (\ref{pint1}). With the help of equations (\ref{eg5}) and (\ref{eg3}), we have
\begin{eqnarray}
(z_{j+1}, t_{j+1}|z_j, t_j)&=&(z_{j+1}|e^{-\frac{i}{\hbar}\epsilon\hat{H}}|z_j)\nonumber\\
&=&\int_{-\infty}^{+\infty}d^{2}p_j~e^{-\frac{\theta}{2\hbar^{2}}\bar{p}_j p_{j}}
e^{i\sqrt{\frac{\theta}{2\hbar^{2}}}\left[p_{j}(\bar{z}_{j+1}-\bar{z}_{j})+\bar{p}_{j}(z_{j+1}-z_{j})\right]}\nonumber\\
&&\times e^{-\frac{i}{\hbar}\epsilon[\frac{\bar{p}_j p_{j}}{2\mu}+\frac{\mu \theta}{4}(\omega_1^2-\omega_2^2)(\bar{z}_{j+1}^2+ z_{j}^2)+\frac{\mu \theta}{4}(\omega_1^2+\omega_2^2)(2\bar{z}_{j+1}z_{j}+1)
-\frac{i\gamma}{2\mu}\sqrt{\frac{\theta}{2}}(p_j \bar{z}_{j+1}-\bar{p}_j z_j)]}+O(\epsilon^2). \nonumber
\label{inftm}
\end{eqnarray}
On substituting the above expression in eq.(\ref{pint1}) and computing the star products, we obtain 
\begin{eqnarray}
(z_f, t_f|z_0, t_0)=&&\lim_{n\rightarrow\infty}\int \prod_{j=1}^{n} (dz_{j}d\bar{z}_{j})
\prod_{j=0}^{n}d^{2}p_{j}\nonumber\\
&&\exp\left(\sum_{j=0}^{n}\left[\frac{i}{\hbar}\sqrt{\frac{\theta}{2}}\left[p_{j}\left\{\left(1+\frac{i\epsilon\gamma}{2\mu}\right)\bar{z}_{j+1}-\bar{z}_{j}\right\}+\bar{p}_{j}\left\{z_{j+1}-\left(1+\frac{i\epsilon\gamma}{2\mu}\right)z_{j}\right\}\right] +\sigma p_{j}\bar{p}_{j}
\right]\right.\nonumber\\
&&\left.~~~~~~~~~~~~~~~~+\frac{\theta}{2\hbar^{2}}\sum_{j=0}^{n-1}p_{j+1}\bar{p}_{j}-\frac{i}{\hbar}\epsilon V(\bar{z}_{j+1},z_{j})\right)
\label{pint3}
\end{eqnarray}
 
where 
\begin{equation}
\sigma=-\left(\frac{i\epsilon}{2\mu\hbar}+\frac{\theta}{2\hbar^{2}}\right) \nonumber 
\end{equation}
and,
\begin{equation}
 ~~V(\bar{z}_{j+1},z_{j})= \frac{\mu \theta}{4}(\omega_1^2-\omega_2^2)(\bar{z}_{j+1}^2+ z_{j}^2)+\frac{\mu \theta}{4}(\omega_1^2+\omega_2^2)(2\bar{z}_{j+1}z_{j}+1)~~\label{sigmas}
\end{equation} 
The momentum integral can be performed easily since it has been brought to the Gaussian form $\exp(\sum_{i,j}p_{i}M_{i,j}\bar{p}_j)$, where $M$ is a 
$N\times N$ ($N=n+1=T/\epsilon$, $T=t_{f}-t_{0}$) dimensional matrix  given by
\begin{eqnarray}
M_{lr}=\sigma\delta_{l,r}+\frac{\theta}{2\hbar^{2}}\delta_{l+1,r}~.
\label{matrix}
\end{eqnarray}
Note that the presence of off-diagonal terms here is purely a NC effect. Further, this matrix M is a typical kind of Toeplitz matrix called as the circulant matrix whose eigenvalues and normalised eigenvectors are given by \cite{toep}
\begin{eqnarray}
\lambda_{k}&=&\sigma+\frac{\theta}{2\hbar^2}e^{2\pi ik/N}\quad;\quad k\in[0,n]\nonumber\\
u_{k}&=&\frac{1}{\sqrt{N}}(1\quad e^{2\pi ik/N}\quad e^{4\pi ik/N}....)^{T}~.
\label{evalues}
\end{eqnarray} 
On executing the momentum integral, we obtain
\begin{eqnarray}
(z_f, t_f|z_0, t_0)=&&\lim_{n\rightarrow\infty}A\int\prod_{j=1}^{n}(dz_{j}d\bar{z}_{j})
\exp\left(-\vec{\partial}_{z_{f}}\vec{\partial}_{\bar{z}_{0}}\right)\nonumber\\
&&\times\exp\left(\frac{\theta}{2\hbar^{2}}\sum_{l=0}^{n}\sum_{r=0}^{n}
\left\{\left(1+\frac{i\epsilon\gamma}{2\mu}\right)\bar{z}_{l+1}-\bar{z}_{l}\right\}M^{-1}_{lr}\left\{z_{r+1}-\left(1+\frac{i\epsilon\gamma}{2\mu}\right)z_{r}\right\}\right)\nonumber\\
&&\times\exp\left(-\frac{i}{\hbar}\epsilon \sum_{j=0}^{n}V(\bar{z}_{j+1},z_{j})\right)
\label{pintegral1}
\end{eqnarray} 
where $A$ is an irrelevant constant coming from momentum integrations on whose explicit form we are not very interested in.
We now make use of the fact that $z_{l}=z(l\epsilon)$
and $z_{l+1}-z_{l}=\epsilon\dot{z}(l\epsilon)+O(\epsilon^{2})$. 

\begin{eqnarray}
(z_f, t_f|z_0, t_0)=&&\lim_{n\rightarrow\infty}A\int\prod_{j=1}^{n}(dz_{j}d\bar{z}_{j})
\exp\left(-\vec{\partial}_{z_{f}}\vec{\partial}_{\bar{z}_{0}}\right)
\exp\left(\frac{\theta\epsilon}{2\hbar^{2}T}\sum_{l,r,k=0}^{n}
\epsilon\left[\dot{\bar{z}}(l\epsilon)+\frac{i\gamma}{2\mu}\bar{z}(l\epsilon)\right]\right.\nonumber\\
&&\left.\times\left[\sigma+\frac{\theta}{2\hbar^{2}}e^{\epsilon\partial_{(l\epsilon)}}\right]^{-1}[e^{2\pi i(l-r)k\epsilon/T}]\times\epsilon\left[\dot{z}(r\epsilon)-\frac{i\gamma}{2\mu}z(r\epsilon)\right]\right)\exp\left(-\frac{i}{\hbar}\epsilon \sum_{j=0}^{n}V(\bar{z}_{j},z_{j})\right)
\nonumber\\
=&&\lim_{n\rightarrow\infty}A\int\prod_{j=1}^{n}(dz_{j}d\bar{z}_{j})\exp\left(-\vec{\partial}_{z_{f}}\vec{\partial}_{\bar{z}_{0}}\right)\nonumber\\
&&\times\exp\left(\frac{\theta}{2\hbar^{2}T}\sum_{l,r,k=0}^{n}
\epsilon\left[\dot{\bar{z}}(l\epsilon)+\frac{i\gamma}{2\mu}\bar{z}(l\epsilon)\right]\left[-\frac{i}{2\mu\hbar}+\frac{\theta}{2\hbar^{2}}\partial_{(l\epsilon)}+O(\epsilon^2)\right]^{-1}\right.\nonumber\\
&&\left.~~~~~~~~~~~~ \times[e^{2\pi i(l-r)k\epsilon/T}]\epsilon\left[\dot{z}(r\epsilon)-\frac{i\gamma}{2\mu}z(r\epsilon)\right]\right)\times\exp\left(-\frac{i}{\hbar}\epsilon \sum_{j=0}^{n}V(\bar{z}_{j},z_{j})\right)
\label{pintegral2}
\end{eqnarray}
On taking the limit $\epsilon\rightarrow 0$ and performing the sum over $k$, we finally have -
\begin{eqnarray}
(z_f, t_f|z_0, t_0)&=&A\exp\left(-\vec{\partial}_{z_{f}}\vec{\partial}_{\bar{z}_{0}}\right)\int_{z(t_0)=z_0}^{z(t_f)=z_f }\mathcal{D}z\mathcal{D}\bar{z}
\exp({\frac{i}{\hbar}S})
\label{pintegral3}
\end{eqnarray}
where the action $S$ is given as follows :
\begin{eqnarray}
S=\int_{t_{0}}^{t_{f}}dt &&\left[\frac{\theta}{2}\left\{\dot{\bar{z}}(t)+\frac{i\gamma}{2\mu}\bar{z}(t)\right\}\left(\frac{1}{2\mu}+\frac{i\theta}{2\hbar}
\partial_{t}\right)^{-1}
\left\{\dot{z}(t)-\frac{i\gamma}{2\mu}z(t)\right\}-\frac{\mu\theta}{2}(\omega_1^2+\omega_2^2){\bar{z}(t)z(t)}\right.\nonumber\\&&~~~~~~~~~~~~~~~~~~~~~~~~~~~~~~~~~~~~~~~~~~~~~~~~~~~~~~~\left.-\frac{\mu\theta}{4}(\omega_1^2-\omega_2^2)({z^2(t)+\bar{z}^2(t)})\right]~~~
\label{action_ncqm}
\end{eqnarray} 

The equation of motion following from the above action is of the following form :
\begin{equation}
\ddot{z}(t)-i\left\{\frac{\gamma}{\mu} -\mu\theta\frac{(\omega_1^2+\omega_2^2)}{2}\right\}\dot{z}(t)+\left\{\frac{\omega_1^2+\omega_2^2}{2}-\frac{\gamma^2}{4\mu^2}\right\}z(t)=-\frac{i\mu\theta}{2}(\omega_1^2-\omega_2^2)\dot{\bar{z}}(t)-\frac{(\omega_1^2-\omega_2^2)}{2}\bar{z}(t)
\end{equation} 

Splitting into real and imaginary parts, we get

\begin{eqnarray}
\ddot{x}_1+\gamma_1\dot{x}_2+\left\{\omega_1^2-\frac{\gamma^2}{4\mu^2}\right\}x_1=0~~~~~~~;~~\gamma_1=\frac{\gamma}{\mu}-\frac{\mu\theta}{\hbar}\omega_2^2.
\label{eff1}
\end{eqnarray} 

and,
\begin{eqnarray}
\ddot{x}_2-\gamma_2\dot{x}_1+\left\{\omega_2^2-\frac{\gamma^2}{4\mu^2}\right\}x_2=0~~~~~~;~~\gamma_2=\frac{\gamma}{\mu}-\frac{\mu\theta}{\hbar}\omega_1^2.
\label{eff2}
\end{eqnarray}
The same equations were obtained in \cite{Sun} for a 2d simple harmonic oscillator ($\gamma=0$) and also obtained in \cite{rohwer} from a different perspective using additional degrees of freedom. 
The limits $\theta,\hbar\rightarrow0$ in such a manner that $\frac{\theta}{\hbar}\rightarrow0$  reproduces the commutative and classical equations of motion as obtained from the Hamiltonian~(\ref{Final}) and are given by ~-

\begin{eqnarray}
\ddot{x}_1+\frac{\gamma}{\mu}\dot{x}_2+\left\{\omega_1^2-\frac{\gamma^2}{4\mu^2}\right\}x_1=0~~.
\end{eqnarray} 

and,
\begin{eqnarray}
\ddot{x}_2-\frac{\gamma}{\mu}\dot{x}_1+\left\{\omega_2^2-\frac{\gamma^2}{4\mu^2}\right\}x_2=0~~.
\end{eqnarray} 
The coupling factors $\gamma_1, \gamma_2$ (\ref{eff1},\ref{eff2}) are modified which is purely due to the presence of the non-commutative parameter $\theta$. Moreover they are anisotropic ($\gamma_1 \not=\gamma_2$) and basically stems from the anisotropic nature of the harmonic potentials in (\ref{Final}) itself.
\\
\\
The characteristic frequencies of this system of equations (\ref{eff1}) and (\ref{eff2}) can now be obtained by first differentiating (\ref{eff1}) w.r.t. time and then using (\ref{eff2}) and then repeating the process for a second time so that we end up with a fourth order differential equation in $x_1$ only to yield a quartic equation in the frequency $\Omega$, which in turn can be solved easily to obtain the following characteristic frequencies :

\begin{equation}
\Omega_\pm = \sqrt{\frac{\nu_1^2+\nu_2^2+\gamma_1\gamma_2}{2} \pm \frac{1}{2} \sqrt{\gamma_1\gamma_2 (2\nu_1^2+2\nu_2^2+\gamma_1\gamma_2)+(\nu_1^2-\nu_2^2)^2}} \\
\label{spect}
\end{equation}
where,

\begin{equation}
\nu_1^2=\omega_1^2-\frac{\gamma^2}{4\mu^2}~~,~~~  \nu_2^2=\omega_2^2-\frac{\gamma^2}{4\mu^2}. \\ 
\label{defi}
\end{equation}

As a special case of our general result, the same frequencies were obtained in \cite{Pina} for a commutative anisotropic oscillator in a magnetic field using canonical quantization.
\subsection{Canonical quantization}
\noindent In this section, we obtain the energy spectrum of our dynamical system by carrying out canonical quantization by making use of the commuting coordinates $\hat{X}_i^c$  (\ref{cv}), rather than  $\hat{X}_i$. Although, they don't correspond to any physical variables, as discussed previously, being commutative they make the process of diagonalisation of the Hamiltonian easier.
We therefore make use of (\ref{cv2}) to recast the Hamiltonian (\ref{hami}) as

\begin{eqnarray}
\hat{H}=\frac{\hat{{P_1}}^2}{2\mu_1}+\frac{\hat{{P_2}}^2}{2\mu_2}+\frac{1}{2}\mu(\omega_1^2\hat{X_1^c}^2 +\omega_2^2\hat{X_2^c}^2)+\frac{\gamma_2}{2} \hat{X_1^c}\hat{P}_2 -\frac{\gamma_1}{2} \hat{X_2^c}\hat{P}_1
\label{hamil1}
\end{eqnarray} 

where
\begin{equation}
\mu_{1}=\frac{\mu}{(1-\frac{\gamma\theta}{2\hbar}+\frac{\mu^2\theta^2\omega_2^2}{4\hbar^2})}~;~~~~~\mu_{2}=\frac{\mu}{(1-\frac{\gamma\theta}{2\hbar}+\frac{\mu^2\theta^2\omega_1^2}{4\hbar^2})}\\
\label{app3}
\end{equation}
and $\gamma_1$ and $\gamma_2$ are as previously defined in (\ref{eff1}) and (\ref{eff2}) respectively.
To diagonalise the hamiltonian we introduce the following canonical transformation :
\begin{equation}
\begin{pmatrix}
            \hat{X}_{1}^c \\
            \hat{X}_{2}^c \\
            \hat{P}_{1}\\
            \hat{P}_{2}
          \end{pmatrix}
  = \begin{pmatrix}
acos{u} & 0 & 0  & \frac{1}{b}sin{u} \\
0 & acos{u} & \frac{1}{b}sin{u} & 0 \\
0 & -bsin{u} & \frac{1}{a}cos{u} & 0 \\
 -bsin{u} & 0 & 0 & \frac{1}{a}cos{u} \
 \end{pmatrix}
 \begin{pmatrix}
            \hat{q}_{1} \\
            \hat{q}_{2} \\
            \hat{\pi}_{1}\\
            \hat{\pi}_{2}
          \end{pmatrix}
\end{equation}
Note the strange nature of this matrix, which essentially has a block-diagonal form, where there is a "rotation" (i.e. modulo $a,b$) in ($\hat{X}_{1}^c,\hat{P}_{2}$) plane and in ($\hat{X}_{2}^c,\hat{P}_{1}$) plane.
In terms of these transformed variables, the Hamiltonian reduces to -
\begin{equation}
\hat{H}=\sigma_1^2 \hat{\pi}_1^2 + \sigma_2^2 \hat{\pi}_2^2 + k_1^2 \hat{q}_1^2 + k_2^2 \hat{q}_2^2 + \lambda_1\hat{q}_1\hat{\pi}_2+ \lambda_2\hat{q}_2\hat{\pi}_1~.
\end{equation}
where, 
\begin{eqnarray}\label{diagp}
k_1^2=\frac{b^2}{2\mu_2}\sin^2{u}+ \frac{\mu\omega_1^2a^2}{2}\cos^2{u}+ \frac{\gamma_1 ab}{2}\sin{2u} \nonumber\\
k_2^2=\frac{b^2}{2\mu_1}\sin^2{u}+ \frac{\mu\omega_2^2a^2}{2}\cos^2{u}- \frac{\gamma_2 ab}{2}\sin{2u} \nonumber\\
\sigma_1^2=\frac{1}{2\mu_1 a^2}\cos^2{u}+ \frac{\mu\omega_2^2}{2b^2}\sin^2{u}+ \frac{\gamma_2}{2ab}\sin{2u} \nonumber\\
\sigma_2^2=\frac{1}{2\mu_2 a^2}\cos^2{u}+ \frac{\mu\omega_1^2}{2b^2}\sin^2{u}- \frac{\gamma_1}{2ab}\sin{2u} \nonumber\\
\lambda_1=-\frac{b}{2\mu_2a}\sin{2u}+\frac{m\omega_1^2a}{2b}\sin{2u}-\gamma_1\cos{2u} \nonumber\\
\lambda_2=-\frac{b}{2\mu_1a}\sin{2u}+\frac{m\omega_2^2a}{2b}\sin{2u}+\gamma_2\cos{2u} 
\end{eqnarray}
A simple inspection shows that the Hamiltonian will be diagonalized if we set $\lambda_1=0=\lambda_2$, in which case the ratio of the parameters $a,b$ and $u$ are determined to be-
\begin{eqnarray}
\frac{a}{b}= \sqrt{\frac{\mu_1\gamma_2+\mu_2\gamma_1}{\mu\mu_1 \mu_2(\gamma_2 \omega_1^2+\gamma_1 \omega_2^2)}} \nonumber \\
and,~~~ \tan(2u)=\sqrt{\frac{4\mu_1 \mu_2(\mu_1\gamma_2+\mu_2\gamma_1)(\gamma_2 \omega_1^2+\gamma_1 \omega_2^2)}{\mu(\mu_2 \omega_1^2-\mu_1 \omega_2^2)^2}}
\end{eqnarray}
The ladder operators are then defined in the usual way as-
\begin{eqnarray}
\hat{a}_{1}=\sqrt{\frac{\sigma_1}{2\hbar k_1}}(\frac{k_1}{\sigma_1} \hat{q}_1+i \hat{\pi}_1)~~;~~~~\hat{a}_{2}=\sqrt{\frac{\sigma_2}{2\hbar k_2}}(\frac{k_2}{\sigma_2} \hat{q}_2+i \hat{\pi}_2)
\label{app4}
\end{eqnarray}
satisfying $[\hat{a}_1 , \hat{a}_{1}^{\dagger}]=1=[\hat{a}_2 , \hat{a}_{2}^{\dagger}]$. With this we finally arrive at the diagonalised form of the Hamiltonian as :
\begin{eqnarray}
\hat{H}=2\hbar k_1\sigma_1(\hat{N}_{1}+\frac{1}{2})+2\hbar k_2\sigma_2(\hat{N}_{2}+\frac{1}{2})~~~~~~~~~~;~\hat{N}_1=\hat{a}_{1}^{\dagger}\hat{a}_1~,~\hat{N}_2=\hat{a}_{2}^{\dagger}\hat{a}_2~.
\label{spectra}
\end{eqnarray}
The energy eigenstates of the Hamiltonian can now be easily labelled by the integer eigenvalues of the number operators $\hat{N}_1, \hat{N}_2$ as $|n_1,n_2)$ satisfying
\begin{equation}
H|n_1,n_2)=\Bigg(\hbar\tilde{\Omega}_1(n_1+\frac{1}{2})+ \hbar\tilde{\Omega}_2(n_2+\frac{1}{2})\Bigg)|n_1,n_2)
\end{equation}
with the characteristic frequencies $\tilde{\Omega}_1=2k_1\sigma_1$ and $\tilde{\Omega}_2=2k_2\sigma_2$\label{spectrumc}.
Remarkably, on substituting the values of $\sigma_1, \sigma_2, k_1, k_2$ from (\ref{diagp}) we find, after a straightforward calculation that these frequencies match exactly with the frequencies (\ref{spect}) obtained from the effective equations of the two modes in the path-integral technique : $\tilde{\Omega}_1=\Omega_+$ and $\tilde{\Omega}_2=\Omega_-$.




Finally note that the expectation values in the coherent state basis of the operator-valued Heisenberg's equations of motion, following from the Hamiltonian (\ref{hami}) are found to be the same as equations (\ref{eff1}) and (\ref{eff2}) obtained from the path integral formalism.\\

\section{Special parametric values~:}
In this pen-ultimate section, we would like to study the possibility of taking vanishing limits\footnote{Here we are not interested in the limit $\eta\rightarrow1$, which corresponds to a constraint system as follows from the structure of the Lagrangian (\ref{G3}) itself resulting in the Dirac bracket  
 \begin{equation}
 [x,y]_{\small{DB}}=\frac{4\gamma}{7\gamma^2 + 8(\epsilon-\omega^2)}
 \end{equation}
 following Dirac's constraint analysis \cite{Dirac}.} of the parameters $\epsilon, \eta$ introduced in (\ref{G1}, \ref{G2}), so that we end up with a pair of Bateman oscillators eventually. Clearly, this limit cannot be executed in any of the intermediate stages as the coordinates $x_1', x_2'$ themselves may fail to make sense. However, if we re-write (\ref{eff1}, \ref{eff2}) in terms of original coordinates $x, y$ (by successively using $T_2^{-1}$ (\ref{t2}) and $T_1^{-1}$ (\ref{t1})) and retain the parameters  $\epsilon, \eta$~ then some remarkable cancellations take place and we end up with completely sensible expressions for (\ref{eff1}, \ref{eff2})-
\begin{eqnarray}
\ddot{x}+ \eta \ddot{y} +(\gamma+\frac{\theta \omega^2}{\hbar}- \frac{\theta\gamma^2}{4\hbar}-\frac{\epsilon\eta \theta}{\hbar})\dot{x}+(\frac{\epsilon\theta}{\hbar}-\frac{\eta\theta\omega^2}{\hbar})\dot{y}+\epsilon y+\omega^2 x=0~~.
\end{eqnarray} 

and,
\begin{eqnarray}
\ddot{y}+ \eta \ddot{x} -(\gamma+\frac{\theta \omega^2}{\hbar}- \frac{\theta\gamma^2}{4\hbar}-\frac{\epsilon\eta \theta}{\hbar})\dot{y}-(\frac{\epsilon\theta}{\hbar}-\frac{\eta\theta\omega^2}{\hbar})\dot{x}+\epsilon x+\omega^2 y=0~~,
\end{eqnarray} 
in the sense that these parameters now occur only in the numerator allowing us to take a smooth limit $\epsilon,\eta \rightarrow0$ which eventually yields -
\begin{eqnarray}\label{effective1}
\ddot{x}+\gamma_R\dot{x}+\omega^2 x=0
\end{eqnarray}
and,
\begin{eqnarray}\label{effective2}
~~~~~~~~~~~~~~~~~~~~~~~\ddot{y}-\gamma_R\dot{y}+\omega^2 y=0~~~; ~ \gamma_R=\gamma+\frac{\theta \omega^2}{\hbar}- \frac{\theta\gamma^2}{4\hbar}.
\end{eqnarray} 

These are clearly in the form of Bateman oscillators (\ref{d1}, \ref{adamped}) except that the damping factor $\gamma$ now takes a renormalized value $\gamma_R$. Both quantum and noncommutative effects are buried here.

Now observe that the characteristic frequencies of the pure Bateman system can be obtained by taking the limits $\epsilon,\eta\rightarrow 0$ in (\ref{spect}). We then, however observe, that although the system becomes isotropic ($\omega_1^2=\omega_2^2=\omega^2-\frac{\gamma^2}{4}$), the mass parameter $\mu$ becomes purely imaginary: $\mu=i$. Not only that, the coupling factors $\gamma_1$ and $\gamma_2$ (\ref{eff1}, \ref{eff2}) for the two modes also become equal and are purely imaginary and we get the characteristic frequencies as-
\begin{equation}
\lambda_\pm^R = i\frac{\gamma_R}{2} \pm \sqrt{\omega^2-\frac{\gamma_R^2}{4}}
\label{spectacle}
\end{equation}
Interestingly, the same characteristic frequencies can be obtained from (\ref{effective1}, \ref{effective2}) directly.
These frequencies are essentially in the same form as $\lambda_\pm$(see page 1) except $\gamma$ is replaced by $\gamma_R$ now.
Again the relevant critical ratio is $R=\frac{4\omega^2}{\gamma_R^2}$. If it exceeds unity ($R>1$) we end up with oscillatory as well as damped and anti-damped solutions corresponding to the two modes x and y respectively of the pure Bateman system. It is worthwhile to mention here that in the commutative limit ($\theta=0$), the oscillatory part of the characteristic frequencies (\ref{spectacle}) matches exactly with the frequency as obtained from 't Hooft's scheme (eqn. (15) of \cite{Vit2}).

Finally we come to the pair of equations (\ref{effective1}) \& (\ref{effective2}) regarding which two important observations can be made at this stage :-\\
\textbf{.} Even if $\gamma=0$ initially, $\gamma_R$ is non-zero: $\gamma_R=\frac{\theta\omega^2}{\hbar}$. This indicates that quantum effects along with NC can induce damping.\\
\textbf{.} On the other hand, if $\frac{\gamma^2}{4}>\omega^2$, we can fine-tune $\theta$, taken as a free parameter, to the following value: $\theta=\frac{\gamma\hbar}{\frac{\gamma^2}{4}-\omega^2}$, so that $\gamma_R=0$.\\
This indicates that an original dissipative theory in commutative space can be mapped to a non-dissipative NC theory, so that we can in a certain sense, trade between noncommutativity and dissipation establishing a duality between these two aspects. In other words, we are able to eliminate the damping or anti-damping solutions and retain only the oscillatory part in a NC space.


\section{Conclusion}
We have shown that a pair of damped and anti-damped oscillators, the so-called Bateman oscillators when placed in an ambient NC space of Moyal type goes over to another Bateman oscillator with renormalized damping factor. We have shown this by adopting a different approach of quantization by augmenting the original Hamiltonian with harmonic oscillator Lagrangian and taking vanishing limits eventually. This was necessary as the original Hamiltonian, in the indirect representation did not have a lower bound. Our analysis in this sense is different from earlier works done in the spirit of 't Hooft's scheme of quantization. Further our quantization was carried out completely by using HS operator formulation in both path integral and canonical quantization schemes. Our final observation is that the renormalized damping can have a non vanishing value even in the absence of "bare" damping. In addition, it could be shown that in certain situation, one can map a dissipative commutative system to a non-dissipative noncommutative system indicating the existence of a duality symmetry between these two aspects. It will be interesting to extend this work in the quantum field theoretic level.
\section*{Acknowledgements}

One of the authors, SKP will like to thank UGC-India for providing financial support in the form of fellowship during the course of this work. We want to thank F.G. Scholtz and Sibasish Ghosh for useful discussions.

\section{Appendix}
Here we provide a brief review of 't~Hooft's scheme \cite{hoo} of quantization of Hamiltonians of
the form  
$H = \sum_{i}p_{i}\, f_{i}(\textbf{q})$, where $f_{i}(\textbf{q})$ are
non--singular functions of the canonical coordinates $q_{i}$.
In this scheme, quantization is achieved only as a consequence of the dissipation of information.
Typically this class of Hamiltonians are not bounded from below.
This might be cured by looking for a time independent positive function $\rho(\textbf{q})$ satisfying $\{\rho,H\} = 0$   so that the Hamiltonian $H$ can be splitted as

$~~~~~~~~~~~~~~~~~~~~~~~~~~~~~~~~~H = H_1 - H_2$~~, with $ H_1 = \frac{1}{4\rho}\left( \rho + H\right)^{2}\;\; ;\;\;
H_2 = \frac{1}{4\rho}\left( \rho -
H\right)^{2}\, ,  \label{1.5}
$

Here $H_1$ and $H_2$ are positively
(semi)definite.
To get the lower bound for the Hamiltonian one thus imposes the constraint condition onto the Hilbert space:
$
H_2\ket{\psi}_{ph} = 0 $ at the operator level, which projects out the states responsible for the
negative part of the spectrum.
Therefore, we have $H\ket{\psi}_{ph} =\rho(\textbf{q})\ket{\psi}_{ph} $.
The Hamilton's equation now reads:
\begin{eqnarray}
\dot{q_{i}}=f_{i}(\textbf{q})=\{q_{i},\rho(\textbf{q})\}
\end{eqnarray}
This clearly indicates a non vanishing Poisson bracket(or the corresponding commutator at the operator level) among the $q_{i}$'s leading to noncommutative structures ensuring that non-vanishing of $\dot{\textbf{q}}$ in the general case.
This has some similarity with the well-known Landau problem of a charged particle moving in a plane and subjected to a constant magnetic field normal to the plane. Here the original commuting coordinate operators $\hat{x}, \hat{y}$: $[\hat{x}, \hat{y}=0]$ when projected to the subspace associated with the lowest Landau level become noncommutative.







\vskip 0.5 cm

\end{document}